%% file: jcd_kepler_v2a_red.tex
\documentclass[mypaper,7pt,twoside]{CoAst}
\usepackage{epsf,graphicx,fancyhdr}
\input{CoAst_layo}

\def\note #1]{{\bf #1]}}

\begin{document}
\sf

\chapterDSSN{Asteroseismology with the {\em Kepler mission\/}}
{J. Christensen-Dalsgaard et al.}
%{J. Christensen-Dalsgaard, T. M. Brown, R. L. Gilliland, H.~Kjeldsen,
%W. J. Borucki \& D. Koch}

\authors{J. Christensen-Dalsgaard$^{1,2}$, 
T. Arentoft$^{1,2}$,
T. M. Brown$^3$,
R. L. Gilliland$^4$, H.~Kjeldsen$^{1,2}$,
W. J. Borucki$^5$ \& D. Koch$^5$}
\Address{$^1$ Institut for Fysik og Astronomi, Aarhus Universitet, 
DK-8000 Aarhus C, Denmark\\
$2$ Danish AsteroSeismology Centre\\
$^3$ Las Cumbres Observatory, 6740B Cortona Dr, Goleta, CA 93117, USA\\
$^4$ Space Telescope Science Institute, Baltimore, MD 20771, USA\\
$^5$ NASA Ames Research Center, Moffett Field, CA 94035, USA}

\noindent
\begin{abstract}
NASA's {\em Kepler mission\/} will fly a photometer based on a
wide-field Schmidt camera with a $0.95\,{\rm m}$ aperture,
staring at a single field continuously for at least 4 years.
Although the mission's principal aim is to locate transiting
extrasolar planets, it will provide an unprecedented opportunity to make
asteroseismic observations on a wide variety of stars. 
Plans are now being developed to exploit this opportunity to the fullest.
%by way of a cooperation
%involving the mission itself, the facilities of the Space Telescope Science
%Institute, modelling and analysis capabilities at the University
%of Aarhus, and proposals from scientists not connected with the mission.
\end{abstract}

% ... has been done by Dziembowski & Goode (1996), Bigot et al. (2000) ....

\section{Introduction}

%\note [JC-D]
%
%\note [Overview of Kepler mission, principal goal of transit observations,
%with follow-up, characterizing planetary systems and search for 
%Earth analogues].
%
The {\em Kepler mission\/} was selected for NASA's discovery programme in 2001,
with a launch now planned for November 2008.
The goal of the mission is to search for extrasolar planetary systems 
with the transit method,
by detecting the slight decrease in the brightness of a star as a planet in 
orbit around it passes in front of the star.
This is probably the most efficient method to detect substantial numbers of 
planets of modest size, and a key goal of the mission is in fact the search
for `Earth analogs', planets of roughly Earth size in year-long orbits 
around solar-like stars.
More generally, planets in the `habitable zone', where conditions are such as
to allow liquid water, are emphasized;
thus the mission is a key component of NASA's Exploration Roadmap.
These goals require very high differential photometric precision and 
observations of a given
field for several planetary orbits, i.e., several years.
Also, to achieve a reasonable probability for the detection of planets a
very large number of stars must be observed, requiring a large field of view
of the photometer.

The requirements for planet-transit detection
also make the {\em Kepler mission\/} very well suited for asteroseismology.
The photometric precision required to study solar-like oscillations is similar
to that needed to detect Earth-size planets,
and the large field ensures that a very substantial number of interesting 
targets will be available, both solar-like pulsators and other types 
of pulsating stars.
Consequently an asteroseismic programme is being established 
within the {\em Kepler\/} project.

Pulsations are found in stars of most masses and essentially all stages of
evolution. The frequencies are determined by the internal sound-speed and
density structure, as well as rotation and possibly effects of magnetic fields,
and the amplitudes and phases are controlled by the energetics and dynamics
of the near-surface layers, including effects of turbulent convection.
Observationally, the frequencies can be determined
with exceedingly high accuracy compared with any other quantity relevant to
the internal properties of the stars. Analysis of the observed frequencies,
including comparison with stellar models, allows determination of the properties
of the stellar interiors and tests of the physics used in the model computation
(e.g., Kjeldsen \& Bedding 2004).

Stars showing oscillations similar to those observed in the Sun are particularly
promising targets for asteroseismology, owing to the large number of generally
well-identified modes that can be observed. Also, the extensive experience from
analyses of solar oscillations can be applied in the analysis of data for these
stars, which have oscillation periods of minutes to hours. Furthermore, the
properties of the oscillations (amplitudes, frequencies, mode lifetimes) show
long-term variations caused by stellar activity. 
%Obtaining long-term time series
%data with Kepler will therefore provide information on stellar activity and
%variations in the magnetic field of a given star, 
%at a level of detail currently only available for the Sun.

Here we give a brief description of the {\em Kepler mission\/} and the
planned asteroseismic investigations.
Further details on the mission were provided by
Basri et al. (2005) and Koch et al. (2007),
%\note [a few printed references],
as well as on the mission web page ({\tt http://kepler.nasa.gov/sci/}).

%\note [Do we have reasonable overview references on the mission?].

%\note [A little on how it fits into the NASA programme, perhaps].

\section{{\em Kepler\/} instrumentation}

%\note [TMB, RLG]
%
%\note [Discuss optics, detector system, briefly.
%Probably include figure showing cross-section, or whatever, 
%of {\em Kepler\/} photometer (in ITAR-clean form!)].
%
The {\em Kepler\/} photometer is a classical Schmidt design
with a 0.95 m diameter corrector passing light to a
1.4 m primary and then on to the focal plane mounted near
instrument centre (see Figure 1).
The focal plane is populated with 42
CCDs with 2200 columns and 1024 rows each that will be read
out through two amplifiers per CCD.  Pixel sizes of 27 $\mu$
will provide full well depths of approximately 1.0 $\times
10^6$ electrons for these backside-illuminated, thinned
and anti-reflection coated devices.  The resulting pixel
scale of 3.98 arcsec results in a large field of view
subtending over 100 square degrees.
The spacecraft is three-axis stabilized with an expected
jitter of less than 1 per cent of the pixel scale.
%jitter smaller than the excellent levels realized for the {\em Hubble
%Space Telescope} when measured on the scale of pixels.

Since tight focus in not required for obtaining optimal
time-series photometry the individual CCD modules are allowed to have
significant focus offsets relative to each other easing
integration of this large focal plane. 
%The result will be
%that some modules provide sharp point spread functions in
%which the large pixels undersample the provided light
%distribution, while other modules will be near critical
%sampling.
Modules with the best focus will have point
spread functions (PSF) with full width at half maximum (FWHM) less than
one pixel resulting in undersampling, while other modules
with larger focus offsets will provide PSFs with FWHM of
about two pixels resulting in critical sampling of the PSF.
On the other hand, focus stability will be tightly constrained.

%\figureDSSN{photometer_cross_sec.eps}
\figureDSSN{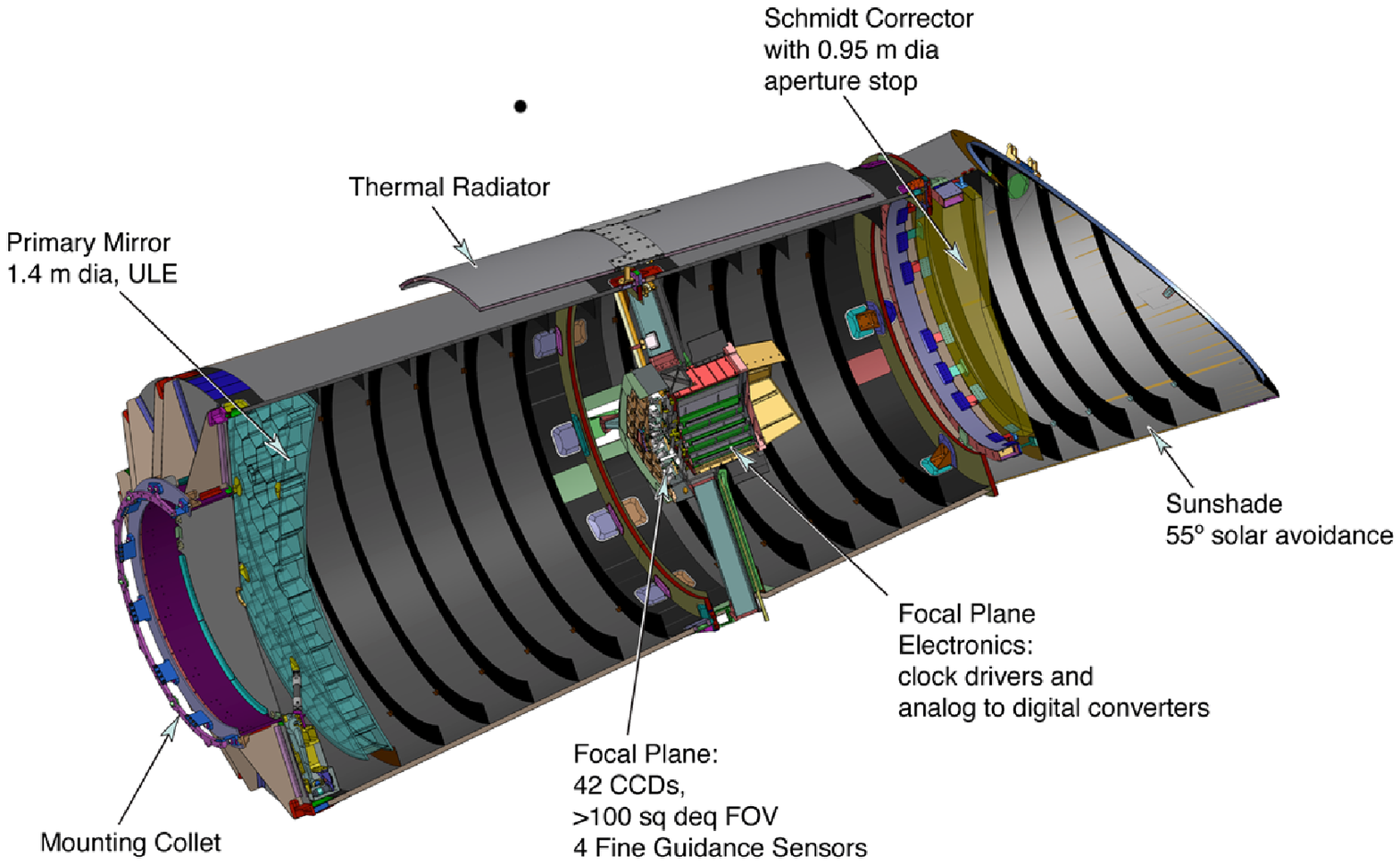}
{%\note [Some caption; could do better for quality of image]
Primary components of the {\em Kepler\/} Photometer
shown in cut-out.  For a higher resolution, colour
version see {\tt http://kepler.nasa.gov/sci/}.
This web site provides a wealth of technical and scientific information
about the mission.
}
{fig:photometer}{!ht}{clip,angle=0,width=80mm}

\section{The {\em Kepler\/} observing programme}

%\note [RLG, TMB]
%
%\note [One field for at least four years. 
%Download postage stamps around target stars (how many?).
%512 targets at one-minute cadence, the rest at 30-min cadence.
%Roll and download gap every three months. (Also likely time for possible 
%change of targets.)].
%
%\note [Here probably also discuss briefly magnitude limits, possibility for
%observing saturated stars].
%
A single field near right ascension 19.4 h and 
declination $44^\circ$ North will be
monitored for the full 4-year mission (with option for a 2-year extension). 
The spacecraft will be in an Earth-trailing
heliocentric orbit, similar to {\em Spitzer\/}.  To keep the
solar arrays illuminated and the focal-plane radiator pointed towards deep 
space
%, located on one side of the spacecraft, illuminated 
%a $90^\circ$ roll will be executed 4 times per year. 
the spacecraft is rotated $90^\circ$ every three months.
%The CCDs are mounted in twos in a $5\times 5$ array
%(in which corners are replaced by $512\times 512$ CCDs
%used for guiding) such that targets will maintain similar
%positions with respect to rows and columns as new detectors
%are rotated into position each quarter.
Figure 2 shows the
CCD coverage superposed on the sky in the Cygnus-Lyra region;
the CCD layout is four-fold symmetric so that the quarterly roll will not
change the sky coverage.
%projected along the Orion arm of the galaxy. 
Transfer of the accumulated data to ground stations,
in the form of small images around each target, will require
body-pointing the high-gain antenna once per month resulting
in data gaps less than one day, in addition to the similar
gaps at the quarterly rolls.

The primary {\em Kepler\/} science searching for transits
of Earth-like planets will be fulfilled by collecting data
on 170,000 stars for the first year, reduced to 100,000
later as high-noise stars are dropped, to accomodate the lower data rates
as the spacecraft drifts away from the Earth.  These targets will
range in magnitude from about 9th to 15th with the design
point being the ability to detect the 85 parts-per-million
(ppm) transits of an Earth analog. % with a $V$=12 G2V host.
The design point is a combined differential photometric
precision of less than 20 ppm in 6.5 hours (half the
length of a central passage of an Earth analog) for a $V$=12 G2V host
when all noise terms are included, 
assuming an intrinsic 10 ppm noise from the solar-like star.
%\note [This has to be related to oscillation detection].
In order to accumulate the 5 $\times 10^9$
electrons at 12th mag without saturating the CCDs, they  will
be read out every 2.5 to 8 seconds (exact value yet to be
set) and accumulated on board into 30-minute sums.  
%To further reduce the data volume only those pixels (ranging
%from 85 at the bright end to 9 at the faint end)
%contributing positively to the signal-to-noise (S/N) are
%telemetered to the ground.

For the extrasolar planet detection, targets that are
dwarfs are strongly preferred over giants;
hence a full ground-based, multi-band photometric screening will be
completed before launch, capable of providing a target list
dominated by F, G and K dwarfs with as many M dwarfs,
to a limit of $V$=16 in this case, as possible.  Due to the
30-minute observing cadence asteroseismology from these
primary observations will be limited to red giants that
have slipped through the screening process (or intentionally
left in), and classical oscillators for which this long
cadence allows Nyquist sampling.

The capability of {\em Kepler\/} to provide also excellent
results for asteroseismology on solar-like stars has been
recognized from the time of initial mission proposals, and a small
complement of 512 targets that can be changed on a
quarterly basis will be followed with 60-second data
accumulations.  
For detailed study of solar-like oscillations the goal should be to reach
a mean photon-noise level in the amplitude spectrum of
1 ppm after three months;
this requires the collection of the $10^{12}$ electrons per month,
which will occur at $V$=11.4.
%This also happens to be approximately the magnitude level
%at which detector saturation is expected per $\sim$3 sec
%integration for CCD modules in better than average focus.
%Since the best asteroseismic constraints are
%expected for targets in which noise levels better than
%1 ppm per month can be obtained, targets of 9th to 11th
%magnitude that are saturated in individual readouts will
%be at the core of this science. 
Stars brighter than this, with photon noise below 1 ppm per month,
are likely the prime targets for asteroseismology.
Such targets are saturated in individual readouts;
however, {\em HST} experience 
%with three different CCD/instrument combinations 
has been that saturated data can support near photon-noise-limited
differential time-series photometry,
%provided that the
%detector gain is adjusted to fully sample the full well
%depth -- as will be required
with a detector set-up such as will be used 
for {\em Kepler\/}.
At $V$=9, usually taken to be the bright limit for
{\em Kepler\/} observations,
%central pixels of the PSF will
%over-saturate by approximately a factor of 10, thus leading
%to bleeding into a comparable number of pixels.
%At 9th magnitude 
the photon-noise limit will be $\sim$70 ppm
per minute, and experience from {\em HST} and simulations for
{\em Kepler\/} suggest that we should be able to do better than
%100 ppm per minute, allowing noise over a one month data
%segment to reach less than 0.5 ppm.
100 ppm per minute, allowing the mean noise level over a three-month data
segment to reach less than 0.5 ppm in the amplitude spectrum.
%In principle, even brighter stars can still provide excellent
%photometry as long as the saturated columns do not extend
%off the CCD resulting in charge loss or, as will more
%commonly be the case, have saturated columns overlap with
%other bright and/or variable stars compromizing the photometry.

Early in the mission the 512 one-minute cadence targets
will be dedicated to those deemed best for asteroseismology.
After the detection of planet candidates from the 170,000
long-cadence targets, many of these
providing high S/N will be switched to the short cadence
to allow refinement of transit shape, timing of transits
for detection of other planets, and also for asteroseismology,
since a prime motivator for the latter
is the exquisite refinement of stellar parameters
(especially radius) thereby obtained.
A substantial number of targets will be reserved for
asteroseismology throughout the mission, however.

\figureDSSN{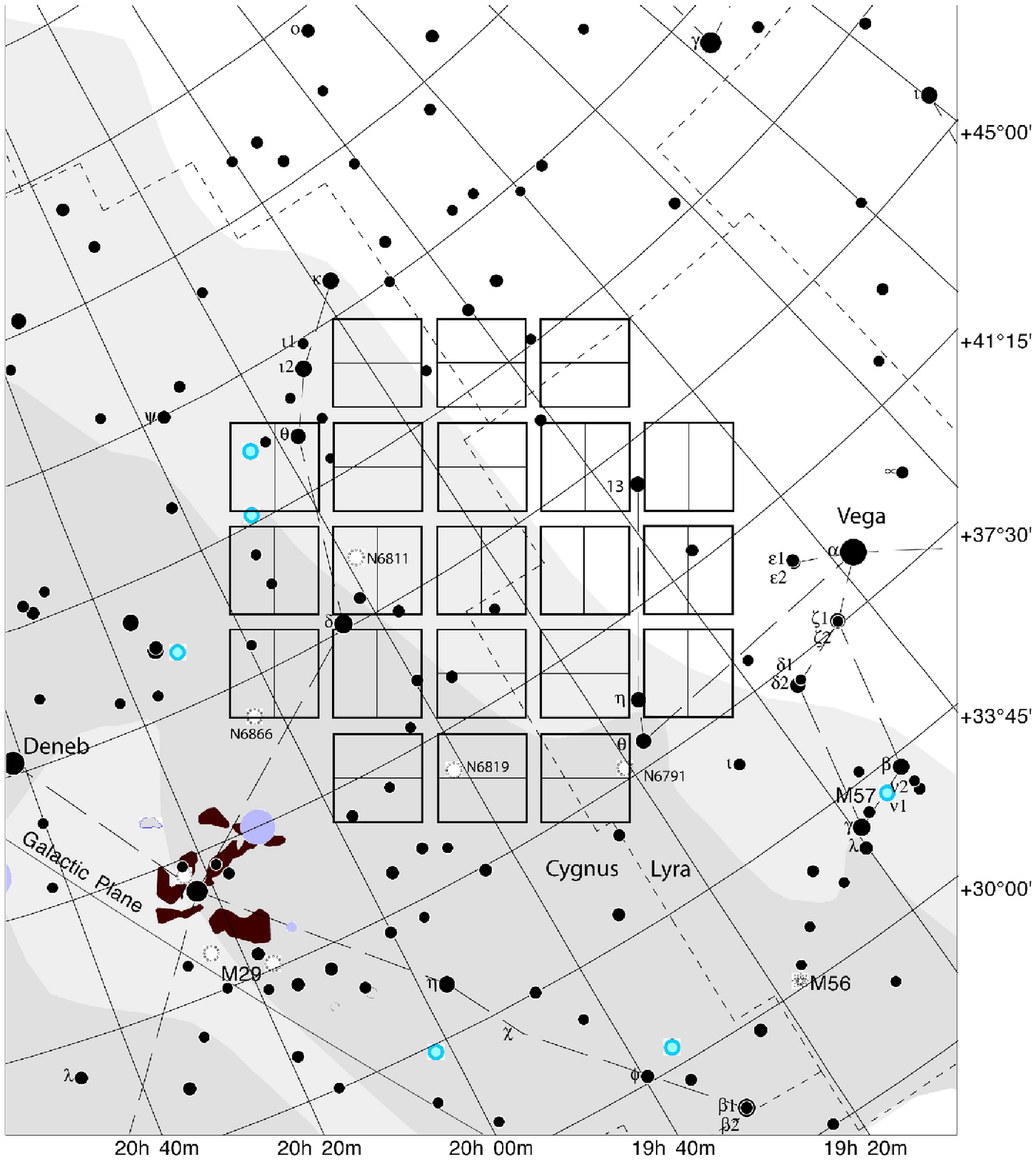}
{%\note [Could do better for quality of image]
Region of galaxy to be monitored with {\em Kepler\/}
showing in detail the layout of the 42 science CCDs.
From {\tt http://kepler.nasa.gov/sci/}.
}
{fig:field}{!ht}{clip,angle=0,width=80mm}

\section{Asteroseismology with {\em Kepler\/}}

%\note [HK]
%
%\note [Obvious point: extremely valuable for all sorts of pulsating stars
%found in the field; need to survey field for interesting cases.
%Note potential of low-cadence observations for classical pulsators, 
%solar-like oscillations in giants].
%
%\note [Here concentrate on solar-like pulsators. 
%Brief discussion of proposed technique to find large and small separations,
%expected magnitude limits. Might be worth showing Arentoft results (HR diagram
%with limiting magnitudes - that actually deserves a separate publication)].
%
%\note [Emphasize potential of observing one star for 4+ years or
%repeatedly, particularly with regards to activity-related frequency changes].
%
The solar-like oscillations are characterized by a great deal of regularity
that relates directly to stellar parameters. This includes in particular the
so-called large and small frequency separations
(e.g. Christensen-Dalsgaard 2004).
Extracting these quantities
from the oscillation signal allows precise determinations of stellar radii
(relative accuracy of 2--3 per cent);
also, ages can be determined with a precision of
better than 5--10 per cent of the total main sequence lifetime,
although the accuracy may be somewhat compromised by uncertainties 
in stellar physics and composition.
We are currently developing techniques for extracting
this information; the large separation can be determined from the power spectrum
of the time-series using cross-correlation and peak comb analysis, and having
obtained that, the small separation can be obtained by a folding of the power
spectrum based on the large separation.

\figureDSSN{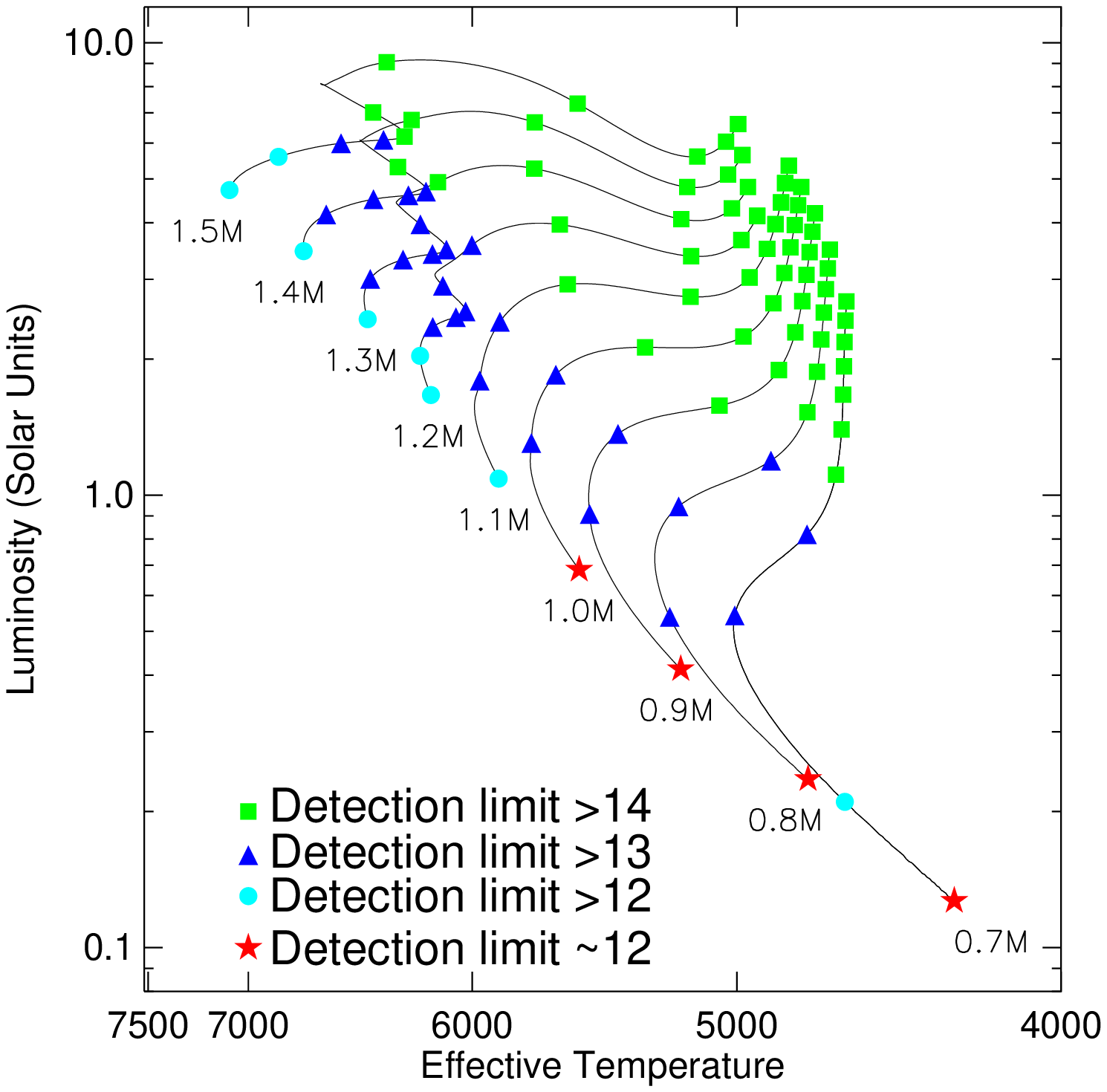}
{
HR diagram of calculated models, with masses in solar units,
indicating the limiting magnitudes to
which the correct large separation could be retrieved from simulations of
one year of Kepler data (see text).
}
{fig:detection}{!ht}{clip,angle=0,width=70mm}

The solar-like oscillations occur in stars across the HR diagram, with
increasing amplitudes and decreasing periods for increasing luminosity
(e.g. Kjeldsen \& Bedding 1995).
%for the relevant scaling relations.
In order to
test our ability to extract stellar parameters using solar-like oscillations,
we calculated oscillation spectra from theoretical stellar models,
and simulated 1-year {\em Kepler\/} time-series including
stochastic excitation of
the oscillations, realistic levels of photon-noise, and granulation. We
calculated time-series for a total of 99 models in the mass-range
0.7--1.5$M_{\odot}$ from the main sequence to the giant branch. For each one,
we added noise corresponding to $V=9-14$ in steps of 0.2 mag, and for each
magnitude value we simulated 10 time-series using different random numbers
for generating the noise.
We then used the analysis briefly discussed above to extract the large
frequency separation to find, for each model, the limiting magnitude to
which we could extract the correct separation in all 10 realisations of
the noise. 
The results are shown in Fig.~\ref{fig:detection}: from one year of
{\em Kepler\/} data we will be able to determine the large separation,
and hence stellar radii, in a very large fraction of the 
relevant stars in the {\em Kepler\/} field observed at the one-minute cadence.
We also expect to be able to determine the small separation in most of the
cases where we could determine the large separation, but this has not yet been
quantified in any detail.

However, for asteroseismology we will be able to go much further. 
Using the {\em Kepler\/} time-series
we will be able to extract the individual oscillation
frequencies, measure amplitudes, phases and mode life-times, and use this
information to interact with theoretical stellar modelling to measure
stellar masses, luminosity, radii, ages, effective temperatures and rotation
for each of the observed stars, as well as test the details of
the physics of the stellar interiors.

We finally note that the time-scale of pulsation varies widely between
different types of stars. For several types of the classical variables (such
as Cepheids), as well as for solar-like oscillations in giant stars, the
pulsation periods are so long that the low-cadence data will be sufficient
for detailed asteroseismic investigations. The long-term, continuous
observations of {\em Kepler\/} will allow the determination of frequencies
to very high precision.
%without the confusing side-bands usually seen in frequency spectra
%of ground-based data due to non-continuous observations.

\section{The {\em Kepler\/} Asteroseismic Investigation (KAI)}

%\note [JC-D]
%
%\note [Briefly on proposed organization; it would help if we had agreement
%on the Agreement before the paper is submitted.
%Note community participation in the KASC].
%
The {\em Kepler\/} Asteroseismic Investigation will be arranged around
the {\em Kepler Asteroseismic Science Operations Centre\/} (KASOC), which
will be established at the Department of Physics and Astronomy, University of
Aarhus.
An agreement is being established to define the details of this
part of the {\em Kepler\/} project.

The relevant {\em Kepler\/} data will be transferred from the 
Data Management Center at Space Telescope Science Institute to KASOC;
the data will be high-pass filtered, or in other ways modified, so
as to contain no information about planet transits.
At the KASOC amplitude spectra will be determined and the frequencies and other
properties of the stellar pulsations will be extracted.
Also, a preliminary asteroseismic analysis will be made to
determine global parameters of the stars, such as radius, mass and age.
Further detailed analyses will be carried out to determine properties
of the stellar interiors and test stellar modelling, particularly for the
relatively bright targets with high signal-to-noise ratio.

The quantity and quality of asteroseismic data expected from {\em Kepler\/}
are overwhelming:
time series extending over months to years for several thousand stars
are expected.
Also, very substantial development of procedures for data analysis and
data interpretation has to take place before the start of the mission, 
and detailed ground-based observations are needed to characterize 
the prime targets of the asteroseismic investigation.
These efforts far exceed the capabilities of KASOC and the directly
involved Co-Investigators of {\em Kepler\/}.
Consequently, we shall establish a {\em Kepler Asteroseismic 
Science Consortium\/} (KASC), with broad community participation,
to help with the preparations and take part in the analysis of the data.
A call will be made early in 2007 for
applications to join the KASC, requesting indication of the contributions
to be made to the project and the planned uses of the data.

\section{Conclusion}

%\note [JC-D]
%
%\note [Obviously a unique possibility for asteroseismology].
%
The {\em Kepler mission\/} promises unique opportunities for asteroseismology,
in terms of the number and variety of stars that can be studied 
with very high differential photometric precision.
This will provide a comprehensive overview of stellar properties across 
a large part of the HR diagram, including information about the 
excitation and damping of the modes,
and detailed information about the internal structure of a substantial 
number of stars.
Also, the long period over which the {\em Kepler\/} field will be
observed offers the possibility of studying frequency variations
associated with possible stellar activity cycles;
thus a parallel investigation of the activity of stars in the {\em Kepler\/}
field through measurement of the H and K indices 
(e.g., Baliunas et al. 1998)
is highly desirable.

{\em Kepler\/} will follow two years after the launch of the 
{\em CoRoT\/} mission 
%\note [reference to Michel paper?]
which shares many of the characteristics of {\em Kepler\/},
including very high photometric precision and observations over relatively
long periods.
Thus a collaboration with the {\em CoRoT\/} asteroseismic project would
be very valuable; 
this could include experience with the optimal analysis of the time series
to determine the oscillation frequencies,
as well as improved information about the expected amplitudes and 
lifetimes of the modes in the potential {\em Kepler\/} targets.

%\note [A little on relation to CoRoT?].
%
%\note [This is probably the point to discuss briefly what the KAI 
%can contribute to the mission as a whole].
%
%The asteroseismic investigations planned on the basis of the {\em Kepler\/}
The asteroseismic investigations based on the {\em Kepler\/}
data will be very valuable for the exo-planet part of the mission.
As demonstrated above, we expect to determine accurate radii for
a substantial fraction of the planet-hosting stars discovered from
planetary transits; this will substantially improve the determination of
the planet radii from the properties of the transits.
Also, in many cases the asteroseismic data will provide estimates of the
age of the star, of obvious value to the understanding of the evolution of
planetary systems.
However, in the present context the main importance of the data is
obviously their great potential value for our understanding of
stellar structure and evolution.

%\acknowledgments{
%%thank you
%}

\References{
Baliunas, S. L., Donahue, R. A., Soon, W., Henry, G. W. 1998,
ASP Conf. Ser. 154, eds R. A. Donajue \& J. A. Bookbinder, p. 153\\
Basri, G., Borucki, W. J. \& Koch, D. 2005,
%[The Kepler mission: A wide-field transit search for terrestrial planets].
New Astronomy Rev. 49, 478 \\ %-- 485.
Christensen-Dalsgaard, J. 2004,
%[Physics of solar-like oscillations].
Solar Phys. 220, 137 \\%-- 168.
Kjeldsen, H., Bedding, T. R. 1995,
%[Amplitudes of stellar oscillations: the implications for asteroseismology].
A\&A 293, 87 \\
Kjeldsen, H., Bedding, T. R. 2004,
ESA SP-559, ed. D. Danesy, p. 101\\
Koch, D., Borucki, W., Basri, G. et al. 2007,
in Proc. IAU Symp. 240, eds W.I. Hartkopf, E.F. Guinan \& P. Harmanec,
Cambridge University Press, in the press \\
}

\end{document}

%% file: CoAst_layo.tex
\renewcommand{\chaptermark}[1]{\markboth{#1}{}}

\newcommand{\kopf}{\small\itshape Comm. in Asteroseismology\\ Vol. 150, 2006}

\newcommand{\authors}[1]{\begin{center}\normalsize\bf\sf #1 \end{center}}

\renewcommand{\author}[1]{\begin{center}\normalsize\bf\sf #1 \end{center}}
\newcommand{\Address}[1]{\begin{center}\small\sf #1 \end{center}}

\renewenvironment{abstract}{\section*{Abstract}\normalsize\sf}{}
\newcommand{\References}[1]{\begin{flushleft}{\large References\\}\vspace*{2mm}\small #1 \end{flushleft}}

\newcommand{\chapterDSSN}[2]{\chapter[\sf\normalsize #1\\ \footnotesize \hspace*{5mm}by #2 \sf\normalsize][]{#1\\}\rhead[\fancyplain{}{\sf\footnotesize \center{#1}}]{\fancyplain{}{\sffamily\thepage}}\lhead[\fancyplain{\kopf}{\sffamily\thepage}]{\fancyplain{\kopf}{\sf\footnotesize \center{#2}}}}

\newcommand{\figureDSSN}[5]{\begin{figure}[#4]
\centering
\includegraphics*[#5]{#1}
\caption{#2}
\label{#3}
\end{figure}}

\renewcommand{\chaptername}{}